\documentclass[journal=jacsat,manuscript=article]{achemso}

\usepackage[version=3]{mhchem} 
\usepackage{color}
\usepackage{graphicx}
\usepackage{subcaption}
\usepackage{hyperref}

\usepackage{xcolor,colortbl}
\usepackage{nicematrix,tikz}
\usepackage[normalem]{ulem}

\newcommand{\mc}[2]{\multicolumn{#1}{c}{#2}}
\definecolor{Gray}{gray}{0.85}
\definecolor{LightPurple}{rgb}{0.86,0.82,1.0}
\definecolor{LightCyan}{rgb}{0.88,1,1}
\definecolor{add}{rgb}{0, 0, 0}
\definecolor{remove}{rgb}{0.7, 0, 0}
\newcommand{\add}[1]{\textcolor{add}{#1}}
\newcommand{\remove}[1]{}

\newcolumntype{a}{>{\columncolor{LightPurple}}c}
\newcolumntype{b}{>{\columncolor{white}}c}

\newcommand\musr{$\mu$SR\ }

\author{Adam Berlie}
\affiliation{ISIS Neutron and Muon Source, STFC Rutherford Appleton Laboratory, Harwell Campus, OX11 0QX, United Kingdom}
\email{adam.berlie@stfc.ac.uk}
\author{Sayani Biswas}
\affiliation{ISIS Neutron and Muon Source, STFC Rutherford Appleton Laboratory, Harwell Campus, OX11 0QX, United Kingdom}
\author{Alex Louat}
\affiliation{ISIS Neutron and Muon Source, STFC Rutherford Appleton Laboratory, Harwell Campus, OX11 0QX, United Kingdom}
\author{Rhea Stewart}
\affiliation{ISIS Neutron and Muon Source, STFC Rutherford Appleton Laboratory, Harwell Campus, OX11 0QX, United Kingdom}
\author{John Wilkinson}
\affiliation{ISIS Neutron and Muon Source, STFC Rutherford Appleton Laboratory, Harwell Campus, OX11 0QX, United Kingdom}

\title[Future perspective of muons; a quantum particle measuring quantum processes]
  {Future perspective of muons; a quantum particle measuring quantum processes}

\abbreviations{IR,NMR,UV}

\begin{document}

%
%
%
%
%

\begin{abstract}
\noindent Although considered a niche technique, muon spectroscopy provides a unique and complementary insight into a range of different materials from hard condensed matter to biological samples and everything in between. In matter, the muon has a mass of \add{$\frac{1}{9}m_p$ or $207m_e$, and is a} local probe of quantum states that \add{can provide a focus on the bulk properties of materials.} While often interpreted in a classical framework, the muon is itself a quantum particle and it is increasingly common for researchers to take account of this when thinking about muon spectroscopy experiments. In this perspective, we focus on the power of using this quantum treatment of muon spectroscopy, which is a key future direction for the technique.  
\end{abstract}


\section{Introduction}
Often at a first glance, the world of Muon Spin Spectroscopy can feel complex and intimidating. For those without a background in physics, the addition of a new particle to our repertoire can feel like a daunting prospect, and bordering closer to the world of particle physics than materials science or chemistry. However, the reality is far from this case: using muons to study materials can provide us with often complementary and in many cases unique information that other experimental techniques either can’t reveal or extract in a complex and indirect way. The hope is that this perspective on the future directions of the muon spin spectroscopy technique provides some food for thought; with readers thinking about how the technique could provide insight into their own research topics as well as to provide a brief amount of information to demystify the use of muons in materials science and chemistry. What we will attempt to convey is the different ways muons can contribute to a variety of problems, but with emphasis on the quantum nature of the muon. It is this quantum behaviour that allows us to directly probe quantum processes within systems, where one is essentially measuring the local Hamiltonian of the muon and relating this back to local/bulk physical properties of the system in question.

There is also a growing trend globally for research to be directed towards the solving and tackling of prominent issues or problems and using curiosity driven research to drive economic growth \cite{IS8,USA2027Budget}. Many of these key strategic areas hinge on our understanding and discovery of new and novel materials; from understanding quantum systems, to energy storage and optimising green energy processes, and enhancing our knowledge of chemical systems. In all of these cases, muons have something to contribute, where in general muons tend to focus on blue skies and lower technology readiness levels.

\section{A Brief Introduction to Muon Spin Spectroscopy}

\add{To date, \musr has had an important impact across a number of areas spanning materials physics and chemistry. For example, muons provide an exceptionally powerful probe of static and dynamic magnetism, which has proved essential in understanding bulk and local magnetic ordering, quantum ground states and ionic motion in a wide range of materials such as antiferromagnets, quantum spin liquids and battery cathode materials. The properties of the muon as a probe in both weak transverse and zero magnetic fields has underpinned the development of the technique within superconductors, where muons are uniquely sensitive to the weak magnetic fields associated with unconventional superconductivity and are a key probe of the underlying superconducting pairing mechanism (for a recent review see \cite{Blundell2025}). The analogy between the muon and a proton ($\mu^+$) or a hydrogen atom (muonium), as discussed later, has opened the technique up to study defect states and chemical physics; both at a molecular level and in how these species react.}

\add{Rather than provide a comprehensive discussion of the history, development and current impact of muon spectroscopy here, we refer the reader to a general review article on the technique \cite{hillier2022muon} and note that the interested reader can find a more detailed treatment in several textbooks that have been recently published.\cite{Muonchemistry, blundell2022muon, amato2024}. Instead, in what follows of this introduction, we provide a brief overview of some key concepts in muon spin spectroscopy, specifically muon spin relaxation, rotation, and resonance ($\mu$SR), to establish the context for the discussions that follow.}

\subsection{Fundamental properties of muons}
\label{sec:Fundamental properties}
Since the discovery of the muon in 1936, when the famous nuclear physicist, Israel “Isidor” Isaac Rabi famously quipped, “Who ordered that?!”, muons have been a subject of constant fascination. \add{With the first muon spin rotation experiment being performed in 1957 \cite{FirstMuSR}}, the groundwork was laid for utilising the muon as a spectroscopic probe of matter that allows the experimentalist to study the interaction of the muon with its surrounding atomic environment when implanted into chemical samples. 

\add{From a particle physics perspective, muons are elementary leptons, but }
\remove{that exist in both positive and negative charged forms, denoted by $\mu^+$ and $\mu^-$ respectively. The use of both will be discussed in later sections.} what is key is that muons are spin-$\frac{1}{2}$ unstable particles with a mean lifetime of a few microseconds, that can be implanted into materials to act as a local probe of both nuclear and electronic magnetism. \add{In the presence of a magnetic field, the spin of the muon undergoes larmor precession, similar to that observed in Nuclear Magnetic Resonance (NMR), but without requiring an equilibrium spin population.} The decay product of the muon contains the information that relates back to the behaviour of the time evolution of the local nuclear and magnetic fields.

Table \ref{properties} summarises the fundamental properties of muons and how these compare to those of protons and electrons. It is the different gyromagnetic ratios that determine the frequency of the precession and the fields/frequency space that one can measure. 

In a typical NMR experiment, \add{the observable nuclear spin polarisation arises from a small thermal population imbalance established in an applied magnetic field and is subsequently manipulated using radio-frequency pulses}, however muon beams are produced with 100\% spin polarisation. 
\remove{The implanted muons are 100\% spin polarised and} \add{This means that} experiments can be performed in zero applied magnetic fields.

\begin{table}[h]
    \centering
    \renewcommand*{\arraystretch}{1.6}
    \caption{The physical properties of muons, electrons and protons.}
    \resizebox{\textwidth}{!}{
    \begin{tabular}{l | a | b | a | b | a | b}
        \hline
        \rowcolor{LightCyan}
        \mc{1}{} & \mc{1}{Spin} & \mc{1}{Charge} & \mc{1}{Mass} & \mc{1}{Magnetic Moment} & \mc{1}{Gyromagnetic Ratio, $\gamma/2\pi$ (MHz T$^{-1}$)} & \mc{1}{Mean Lifetime (\textmu s)}\\[5pt]
        \hline
        \hline
        electron & $\frac{1}{2}$ & $-e$ & $m_e$ & $647\mu_p$ & 28025 & $\infty$\\[5pt]
        muon & $\frac{1}{2}$ & $\pm e$ & $207m_e$($=0.11m_p$) & $3.18\mu_p$ & 135.53 & $2.2$\\[5pt]
        proton & $\frac{1}{2}$ & $+e$ & $1836 m_e$($=m_p$) & $\mu_p$ & 42.576 & $\infty$\\[5pt] 
        \hline
    \end{tabular} }
    \label{properties}
\end{table}

\add{Muons exist in both positive and negative charged forms, denoted by $\mu^+$ and $\mu^-$ respectively.}
When implanted into materials, $\mu^+$ can be thought of as a light proton and may stop in regions of negative electron density. Whereas $\mu ^{-}$ behaves similarly to a heavy electron and typically comes to rest close to an atomic nucleus. In both cases, the muon will respond to its local environment, where over many muon lifetimes, one can build up the time evolution of the polarisation, relating this back to the behaviour of the sample in question. This is the essence of a \musr experiment.

Whilst the majority of muon spectroscopy experiments involve implanting $\mu^+$ into materials, questions can arise from the motion or diffusion of the light $\mu^+$, that is held in place by electrostatic or electronic charge. In this case negative muon ($\mu^-$) experiments are of particular advantage where it is important that the muon is stationary, for example in experiments that aim to measure the diffusion of ions through a battery cathode material \cite{Sugiyama2018}. $\mu ^{-}$ can also be used to perform negative muon elemental analysis experiments that allow the elemental composition of materials to be determined. This method of elemental mapping offers several advantages over more traditional techniques and is discussed in more detail in the \nameref{negativemuon} section, although the compromise for these two experiments is they often require longer acquisition times.

\subsection{$\mu^+$ States in Matter}
\label{sec:muon states}
\begin{figure}
    \centering
    \includegraphics[trim=20 60 20 60,clip,width=\textwidth]{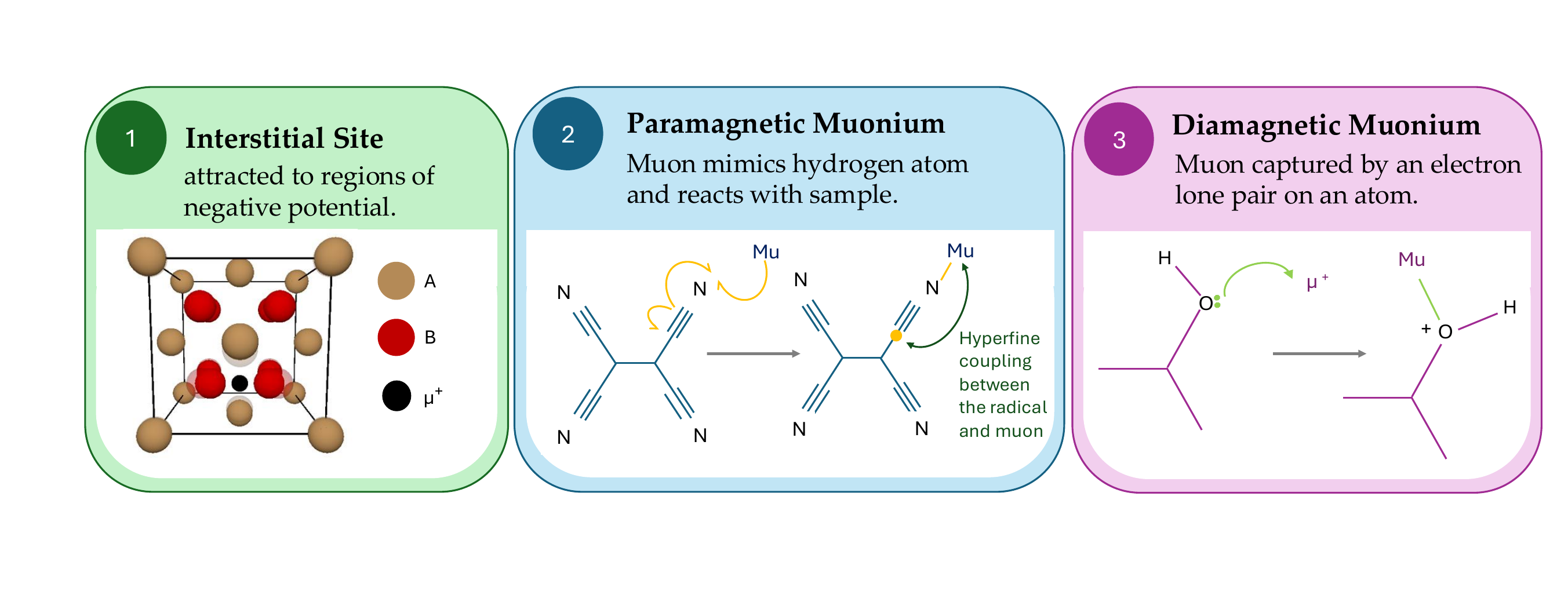}
    \caption{A schematic diagram showing the different muon states that can form in materials 1) a bare muon at an interstitial site; 2) diamagnetic muonium; and 3) paramagnetic muonium.}
    \label{fig:muon_states}
\end{figure}

Following implantation into a material, positive muons rapidly thermalise through a series of complex inelastic processes into one of several different final states; a bare interstitial muon, a chemically bonded diamagnetic muon or muonium state or paramagnetic muonium. Distinguishing between the three states is important and they are illustrated in Figure \ref{fig:muon_states}. 

Muonium (Mu) is a $\mu ^{+}$--$e^{-}$ bound state that can be considered a light isotope of hydrogen. It has a Bohr radius $\approx 1.004a_{\rm H}$, where $a_{\rm H}$ is the Bohr radius of hydrogen, and a reduced mass $m_{\rm Mu}\approx \frac{1}{9}m_{\rm H}$.

In the simplest case, the muon remains as a bare positively charged particle occupying an interstitial site within the crystal lattice, where it may induce a local lattice distortion. The muon interacts with nearby nuclear and electronic magnetic moments through dipolar interactions that decay as $1/r^3$ (see Section \nameref{sec:quantum_probe}). In molecular systems, the muon may instead form a dative bond with a lone electron pair, analogous to proton binding, or muonium may react with a molecular radical to produce a diamagnetic singlet state. In both cases, the local environment is probed through dipolar and quadrupolar interactions. Alternatively, if paramagnetic muonium forms, it behaves similarly to an isolated hydrogen atom and can participate directly in chemical reactions. A common example is addition across double or triple chemical bonds, producing a paramagnetic radical. There is then a hyperfine interaction between the muon, the radical and surrounding nuclei.

Often one can make a good guess where muons will locate or react from simply looking at the crystal or molecular structure and applying some basic chemical knowledge. However computational techniques offer more precision in calculating the stopping sites
\cite{Blundell2025Electronic}. As a general rule of thumb, in metallic systems most muons remain free, whereas in insulators there is a higher probability of muonium formation.

\subsection{Complementarity}

It is important to recognise that as we take a more solutions based approach to chemical and physical problems, we understand how to apply the right technique to solve the problem at hand. Because the muon acts as a local magnetic probe while simultaneously sampling the bulk of the material, $\mu$SR provides information that is often complementary to other experimental methods.

\begin{figure}[h]
    \centering
    \includegraphics[width=1\linewidth]{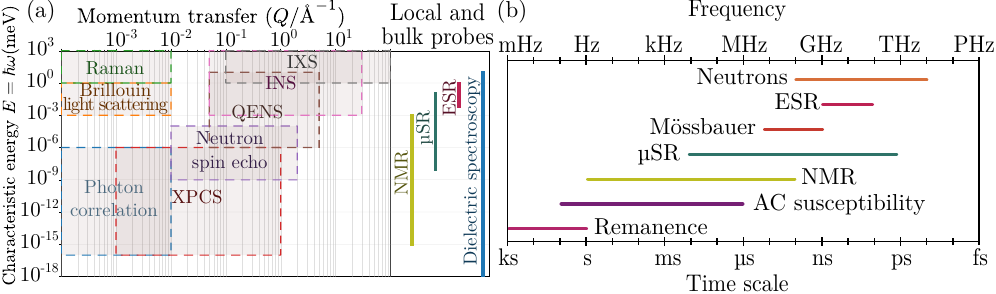}
    \caption{\add{Comparison of the dynamical ranges accessed by selected spectroscopic and scattering techniques.}(a) Approximate \add{characteristic energy scales} \remove{experimental windows of common spectroscopic and scattering techniques for probing dynamics as a function of energy transfer}, $E=\hbar\omega$, and, \add{for scattering techniques,} momentum transfer \add{ranges. Local and bulk probes are shown on the right using an equivalent energy scale derived from their characteristic
frequencies.}\remove{, $Q$. The diagram highlights the complementary ranges covered by Raman and Brillouin light scattering, photon correlation spectroscopy, X-ray photon correlation spectroscopy (XPCS), neutron spin echo, quasielastic neutron scattering (QENS), inelastic neutron scattering (INS), inelastic X-ray scattering (IXS), nuclear magnetic resonance (NMR), electron spin resonance (ESR), muon spin rotation/relaxation ($\mu$SR), and dielectric spectroscopy, distinguishing between $Q$-resolved and non-$Q$-resolved methods.} (b) \add{Approximate}\remove{Characteristic} frequency and corresponding time-scale \add{windows}\remove{ranges} accessible to \remove{selected experimental probes of dynamics. Remanence}\add{remanence measurements}, AC susceptibility, nuclear magnetic resonance (NMR), muon spin rotation/relaxation ($\mu$SR), M\"ossbauer spectroscopy\add{, electron spin resonance
(ESR)}, and neutron \add{spectroscopy}\remove{-based techniques span complementary windows from slow macroscopic relaxation processes to ultrafast microscopic fluctuations}\add{The displayed ranges are schematic and depend on
the instrument, applied field, sample, and experimental protocol. XPCS denotes X-ray photon correlation spectroscopy, QENS quasielastic neutron scattering, INS inelastic neutron scattering, and IXS inelastic X-ray scattering}.}
    \label{Timescales}
\end{figure}

Like all spectroscopic techniques, $\mu$SR is sensitive to a characteristic range of frequencies and timescales. Figure \ref{Timescales}(a) illustrates where $\mu$SR lies within the broader landscape of spectroscopic and scattering methods, highlighting the unique frequency window that it occupies and its complementary relationship with established techniques.

Among the resonance techniques, perhaps the greatest parallels can be drawn between NMR and Electron Spin Resonance (ESR), where much of the underlying theory between the three techniques is extremely similar. NMR and ESR rely on the application of an external magnetic field to lift the degeneracy of nuclear or electronic spin states through Zeeman splitting, after which radio frequency or microwave excitation perturbs the spin populations away from equilibrium. The subsequent relaxation back to equilibrium is characterised through the spin-lattice ($T_1$) and spin-spin ($T_2$) relaxation times. In both cases the spin probes are the nuclei or electrons, respectively.

\add{Like NMR, \musr is a local probe: each implanted muon responds to the
magnetic and electronic environment at its stopping site, while the
ensemble signal samples the bulk of the material. Unlike neutron or X-ray
scattering, \musr does not directly map the response as a function of
momentum transfer. The relationship between the local response and
reciprocal-space correlations is outlined in the Supporting Information.}

\remove{In terms of the length scale over which one measures, $\mu$SR can be considered  similar to NMR where one probes local environments with no momentum transfer. This is different to techniques such as neutron or X-ray scattering, which simultaneously provide information on both energy and momentum transfer. Although each muon only probes its local environment over a radius of approximately 2 nm through dipolar coupling (interactions decrease as $1/r^3$) or hyperfine interactions, the overall muon ensemble provides a measurement averaging over the bulk phase. This combines a local sensitivity with a probe of the bulk behaviour.}

This local selectivity is often advantageous. Because the muon occupies well-defined stopping sites, it can selectively probe the dominant phase while remaining relatively insensitive to minor impurity phases. If the stopping sites are understood, the muon effectively acts as an isotopic label for specific regions of the structure.

Figure \ref{Timescales}(b) shows where the muon technique lies in terms of the timescales \remove{and length scales} that other common techniques measure. Of particular note is that the frequencies (or timescales) that \musr \add{probes}  a gap between AC susceptibility, NMR and neutron scattering. This allows for the MHz frequency range to be probed and behaviours can be traced through a range of different timescales -- something that is extremely valuable when considering systems whose dynamics dominate the behaviour of interest.

\section{The muon as a quantum probe embedded in a quantum system}
\label{sec:quantum_probe}

\begin{figure}
    \centering
    \includegraphics[width=1\linewidth]{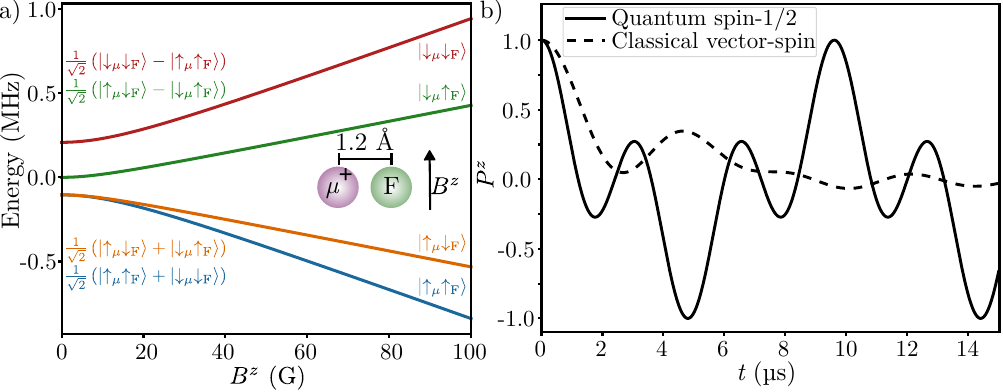}
    \caption{a) Energy levels of the F–$\mu$ pair as a function of the magnetic field $B_z$, with the F–$\mu$ bond oriented along the $x$-axis. Only the dipolar coupling and an external magnetic field applied along the $z$-axis are included. The inset shows the geometry considered. b) Zero-field (ZF) muon polarisation function for unpolarised initial fluorine spins, calculated using quantum spin-1/2 dipolar dynamics and classical vector-spin dipolar dynamics.}
    \label{fig:muon_fluorine_polarisation}
\end{figure}

It is often convenient to introduce \musr in simple classical language: 
the muon is implanted into a material, experiences a local magnetic field, 
and its spin precesses or relaxes in response to that field \add{(similar 
to NMR and ESR, where the measured field is the vector sum of the applied
field and that intrinsic to the sample)}. 
Indeed this \add{classical picture is exactly how the technique was initially 
understood, for example using the result of Kubo and Toyabe's 1966 thought experiment~\cite{kubotoyabe}
to analyse muons implanted in copper \cite{hayano1979zero}.
In 1983 Celio and Meier provided a new interpretation \cite{Celio1983}: 
they realised that the muon was actually 
a quantum probe embedded in a quantum mechanical system, and showed
that treating the muon in this way with interactions which are beyond simple
 Zeeman splitting 
leads to the possibility of an accurate quantitative 
analysis of diffusion processes within copper. Therefore, while it} 
\remove{is how the technique was introduced in the 
\nameref{sec:Fundamental properties} section of this perspectives article. 
This picture} is useful (and in many routine cases sufficient) \add{to think 
of \musr in the classical framework}, \remove{however} it
is ultimately incomplete: the muon is not simply a passive magnetometer 
placed inside a sample \cite{blundell2022muon,hillier2022muon}, but rather 
a quantum particle embedded within a quantum mechanical environment, 
and the signal measured in a \musr experiment is generated by the 
time evolution of this coupled quantum system. 

Therefore, the physical information contained in a \musr experiment 
is ultimately determined by the Hamiltonian that governs the muon 
and its surroundings. In the simplest case, this may indeed reduce to the 
Zeeman interaction between the muon spin $\mathbf{S}_{\mu}$ and a local magnetic field
$\mathbf{B}_{\rm loc}$,
\begin{equation}
\mathcal{H}_{\rm Zeeman}=-\gamma_{\mu}\mathbf{S}_{\mu}\cdot\mathbf{B}_{\rm loc},
\end{equation}
so that the measured precession frequency can be interpreted directly in terms 
of the local field. However, \add{there has been considerable work over the 
past few decades into utilizing other Hamiltonians to probe quantum mechanical 
processes inside materials: for example, in 1986 the dipole-dipole 
interactions was used to probe muon-induced distortions 
in fluorides \cite{brewer1986observation}, which was later extended to a range
of other materials \cite{noakes1993the}, and these results 
were extended further to model
systems to quantify quantum mechanical 
decoherence\cite{wilkinson2020information}. There are also other Hamiltonians
can govern the interactions of the muon, including} 
 \remove{in many scientifically important cases the relevant 
Hamiltonian contains more than this single term: the muon may couple through dipolar 
interactions to nearby nuclear or electronic moments 
\cite{brewer1986observation,noakes1993the,wilkinson2020information},} 
through hyperfine interactions to the local electronic spin density 
\cite{Estle1985,Boekema1984,Onuorah2018}, through quadrupolar terms indirectly 
via neighbouring nuclei, or through time-dependent interactions associated 
with spin fluctuations, charge motion, diffusion, or chemical 
dynamics \cite{hayano1979zero,mcclelland2020muon,Ito2024}. The F--$\mu$ example shown in 
Figure~\ref{fig:muon_fluorine_polarisation} provides a useful illustration of this 
point \cite{brewer1986observation,wilkinson2020information}: in this case, 
the muon is coupled to its nearest-neighbour fluorine nucleus via the dipole-dipole
interaction. Even for this small spin cluster, the coupled energy levels and 
the resulting polarisation function is determined by the quantum Hamiltonian, 
and this differs substantially from a classical vector-spin description. 
Therefore it is important to note that experiments rarely measure 
`a field' in isolation, but the consequence of quantum evolution under the local 
Hamiltonian of the muon--matter system.

In this language, the measured muon polarisation is the expectation value of a muon spin operator after time evolution under the Hamiltonian \cite{blundell2022muon},
\begin{equation}
P^z_{\mu}(t)={\rm Tr}\left[e^{i\mathcal{H}_0t}\rho_0e^{-i\mathcal{H}_0t}\sigma_{\mu}^{z}\right].
\label{timeevolution}
\end{equation}
Here $\rho_0$ describes the density matrix of the initial state of the muon and its 
local environment, and $\mathcal{H}_0$ contains the interactions that entangle, dephase, or
relax the muon spin. The experimental asymmetry (which is directly proportional to 
$P_\mu(t)$) is therefore a direct time-domain readout 
of the quantum interactions governed by $\mathcal{H}_0$. This is one of the reasons why \musr 
is so powerful: the observable is local, microscopic, and dynamical, yet it is measured from 
an ensemble of muons implanted throughout the bulk of the material.

Thinking in terms of the Hamiltonian also clarifies why different classes of \musr experiment 
are sensitive to different physical processes and behaviours. Typical Hamiltonians found in \musr are listed in 
Table~\ref{hamiltonians}. For example, in a magnetically ordered material, the dominant term may 
be the quasi-static coupling of the muon spin to an internal field via the Zeeman Hamiltonian, 
giving coherent oscillations or a broadened field distribution. In a fluctuating magnet, the important 
quantity is instead the spectral density of the local field fluctuations at frequencies \add{sampled by } \remove{relevant to} the 
muon spin evolution\add{ \cite{mcmullen1978}. A more technical discussion is
provided in the Supporting Information.}.

In molecular and chemical systems, muonium formation or muonated radical states 
introduce hyperfine couplings ($A$) that encode information about electronic structure and reactivity. 
These examples look very different experimentally, but they share the same 
underlying principle: the physics is determined by the terms that enter the local muonic Hamiltonian.

\begin{table}[h]
    \centering
    \renewcommand*{\arraystretch}{1.6}
    \resizebox{\textwidth}{!}{
    \begin{tabular}{b | a | a }
        \hline
        \rowcolor{LightCyan} 
        \mc{1}{Name} & \mc{1}{Interaction Hamiltonian} & \mc{1}{Typical systems}\\[5pt]
        \hline
        \hline
        Static Zeeman    &  $\mathcal{H}_{\rm Zeeman} = -\gamma_{\mu}\mathbf{S}_\mu\cdot \mathbf{B}_{\mu}$
            & Statically ordered magnets\cite{Maeter2009,Blundell1997}, 
                TF-\musr\cite{Frandsen2016, Solt1996}\\[5pt]
        Dynamic Zeeman   &  $\mathcal{H}_{\rm Zeeman}(\tau) = -\gamma_{\mu}\mathbf{S}_\mu\cdot \mathbf{B}_{\mu}(\tau)  $
            & Spin liquids\cite{Pratt2022}, frustrated magnets\cite{Barth1989}\\[5pt]
        Dipole-Dipole  &  $\mathcal{H}_{\rm Dip} = \frac{\mu_0}{4\pi}\sum_{ij}\frac{\hbar\gamma_i\gamma_j}{r_{ij}^3}
        \big(\mathbf{S}_i\cdot\mathbf{S}_j-3(\mathbf{S}_i\cdot\mathbf{\hat r}_{ij} )(\mathbf{S}_i\cdot\mathbf{\hat r}_{ij})\big)$
            & Muonic entanglement\cite{brewer1986observation,Lancaster2007,wilkinson2020information}, ionic conductors\cite{mcclelland2020muon} \\[5pt]
        Quadrupolar  & 
            $\mathcal{H}_{\rm Q} = \sum_i \frac{eQ}{\hbar2I_i(2I_i-1)}\sum_{\alpha\beta}V^{\alpha\beta}_i\mathbf{S}^\alpha_i\mathbf{S}^\alpha_j$ 
            & As above, when $I>\frac{1}{2}$ nuclei are involved\cite{wilkinson2020information,Bonfa2022,Berlie2022}\\[5pt] 
        Hyperfine &  $\mathcal{H}_{\rm HF} = A \mathbf{S}_\mu\cdot\mathbf{S}_e$ 
            & Molecular systems\cite{mckenzie2013hyperfine,Dehn2021}, Mu$^0$ in semiconductors\cite{Estle1985,yokoyama2017photoexcited}\\[5pt]
        \hline
    \end{tabular}}
 \caption{Typical muon interaction Hamiltonians. \add{For dynamic local fields,
the relaxation is governed by temporal correlations of the field at the
muon site \cite{mcmullen1978}.}\remove{Note in the dynamic case, the muon polarisation function is obtained by integrating over $\tau$ which results in the relaxation rate being proportional to the field-field correlation function}}
    \label{hamiltonians}
\end{table}

This perspective is especially important because the muon is part of the system 
it probes. As discussed in the \nameref{sec:muon states} section, the implanted $\mu^+$ may 
occupy a particular crystallographic site \cite{Bernardini2013,Moeller2013,Blundell2023} 
and distort its local environment, form a bond, capture an electron to 
produce muonium, or act as an analogue of 
a light hydrogen isotope. These effects are sometimes viewed as complications,
\remove{and they can be.} however, they are also part of what makes the technique distinctive: 
the muon can act as a deliberately introduced, infinitely dilute quantum defect whose 
spin evolution reports on how that defect couples to the surrounding material. Rather 
than treating the muon perturbation only as something to be removed or corrected for, 
future developments in \musr should increasingly exploit it as a controlled way of accessing 
local quantum interactions.

A useful analogy is with other magnetic resonance techniques, where the measured resonance 
frequencies, relaxation rates, and transition probabilities are interpreted through a spin 
Hamiltonian. The difference in \musr is that the spin probe is implanted initially highly 
polarised, which allows one to access local quantum dynamics without the need to create 
a thermally polarised spin population in the sample. In more advanced forms of the technique, 
such as RF-\musr or microwave control of muonium, the Hamiltonian is not only observed but 
actively driven, allowing selected transitions, coherence transfer, or coupled spin 
states to be manipulated.

The future of \musr therefore lies partly in moving from phenomenological relaxation 
functions towards Hamiltonian-based interpretation. Standard models such as static or 
dynamic Kubo-Toyabe functions\cite{kubotoyabe} remain invaluable, but they often make simplifying 
assumptions: that the muon samples a Gaussian field distribution, that the surrounding 
spins are classical and unpolarised, and that the muon does not significantly affect the 
dynamics being measured. As computational methods, electronic structure calculations, 
quantum spin simulations, and data analysis tools improve, it becomes increasingly realistic 
to calculate the muon polarisation function from microscopic Hamiltonians \cite{Gomilsek2023,Onuorah2019,Goli2022,Deng2023,McArdle2021}. This would allow \musr spectra 
to be connected more directly to stopping sites, local bonding, nuclear-spin networks, 
electronic charge density, and dynamical processes.

Viewed in this way, the muon is not merely an unusual experimental probe. 
It is a quantum particle placed inside a quantum material or chemical environment, and the 
experiment follows the way in which its spin state evolves under the local Hamiltonian. 
Interpreting \musr through this framework helps unify the diverse applications of the 
technique, from magnetism and superconductivity to diffusion, charge order, defects, and chemical 
reactivity. It also provides a natural route for the field to develop: by treating the muon not 
as a passive observer of materials, but as an embedded quantum probe whose dynamics reveal 
the microscopic interactions that govern material behaviour.

\add{The above discussion focuses on idealised systems where the interactions are well defined and the spin of the muon couples to a single nucleus or electron. As one introduces additional degrees of freedom, as is the case for chemical systems, the interaction between the muon and its local environment becomes more complex. However, as summarised below, the muon and muonium remain powerful probes of quantum chemical states. Realising their full potential will require a more comprehensive understanding of how these probes interact with and report on their local chemical environments.}

\section{Quantum Chemistry: $\mu^+$ and muonium}
\label{sec:chemistry}
The electronic structure of muonium closely resembles that of atomic hydrogen, making it a light isotope of hydrogen (Table \ref{isotope}). This unique property enables muonium to act as a hydrogen analogue in chemical systems, while the intrinsic 100\% spin polarisation of implanted muons makes \musr a powerful technique for probing chemical reactions and local electronic environments.
Unlike atomic hydrogen, which is often difficult and expensive to produce and manipulate experimentally, muons can be readily implanted into gases, liquids and solids using a wide range of sample environments. A recently published textbook by Fleming, McKenzie and Percival provides an excellent review of muon interactions with chemical systems \cite{Muonchemistry}. Consequently, muonium offers a convenient means of introducing a hydrogen-like probe into materials without the experimental challenges associated with generating atomic hydrogen.

\begin{table}[h]
    \centering
    \renewcommand*{\arraystretch}{1.6}
    \begin{tabular}{b | a }
        \hline
        \rowcolor{LightCyan} 
        \mc{1}{Isotope} & \mc{1}{Relative Mass}\\[5pt]
        \hline
        \hline
        Mu & 1/9\\[5pt]
        $^1$H & 1\\[5pt]
        $^2$H & 2\\[5pt]
        $^3$H & 3\\[5pt]
        \hline
    \end{tabular}
 \caption{Relative mass of muonium to hydrogen isotopes}
    \label{isotope}
\end{table}

Although muonium exhibits chemistry that is broadly analogous to hydrogen, its much lower mass gives rise to important quantum mechanical differences. The reduced mass increases its zero-point energy, leading to greater quantum delocalisation and an enhanced probability of tunnelling compared with a proton. As a consequence, muonium samples the potential energy surface differently from hydrogen and can access reaction pathways that are more sensitive to nuclear quantum effects. As interest in the quantum mechanical nature of chemical reactions continues to grow \cite{Baiardi2021}, these differences provide unique opportunities to investigate the role of quantum effects in chemical reactivity. Additionally, it may also provide more understanding of the zero-point motion and tunnelling of the muon within a range of materials, where intrinsic motion of the muon can have effects on the processes measured \cite{Onuorah2019,Hotz2026}.

The chemistry of muonium has been studied for several decades but remains relatively underutilised. Because it behaves as a light isotope of hydrogen, muonium can be implanted directly into materials where it participates in chemical reactions in much the same way as atomic hydrogen. This enables the formation of transient reaction intermediates and final products to be studied in situ. A particularly well-established example is the addition of muonium across carbon--carbon double bonds, leading to the formation of paramagnetic radical species whose electronic structure and dynamics can subsequently be investigated.

The resonance capabilities of \musr provide an additional level of chemical sensitivity. Hyperfine interactions between the unpaired electron of muonium and neighbouring nuclei produce characteristic level-crossing resonances, from which hyperfine coupling constants can be determined directly. Furthermore, applying an external magnetic field allows interactions to be probed at specific frequencies according to $\nu=\gamma B$. For diamagnetic muons the relevant gyromagnetic ratio ($\gamma$) is that of the muon, $\gamma_\mu$, whereas for paramagnetic muonium the much larger electron gyromagnetic ratio, $\gamma_e$, dominates the spin dynamics. Electron spin flips therefore induce rapid dephasing of the muon spin, allowing measurements of the muonium spin-lattice relaxation time ($T_1$) to probe dynamical processes occurring on significantly shorter timescales than are accessible using diamagnetic $\mu^+$ states.

More recently, these unique quantum mechanical properties have been exploited to investigate chemically and industrially relevant systems. By combining the hydrogen-like reactivity of muonium with the sensitivity of $\mu$SR, it has become possible to identify transient radical intermediates and characterise local electronic environments \cite{Muonchemistry}. A recent example of this has been to probe complex reaction mechanisms in heterogeneous catalysts \cite{Berlie2026}. These capabilities demonstrate the growing role of muon spectroscopy as a complementary tool for understanding chemical reactivity at the atomic scale.

\add{In many of these systems, the magnetic interactions experienced by the muon are large, highly anisotropic, or strongly dependent on the local chemical environment through hyperfine and dipolar couplings. While conventional $\mu$SR measurements provide valuable information through passive observation of the spin evolution, even greater insight can be obtained by actively perturbing the spin Hamiltonian. The application of external stimuli, such as oscillating magnetic fields, enables specific spin transitions to be driven and controlled, providing direct access to otherwise inaccessible aspects of the local quantum state and dynamics. The development of these techniques has significantly expanded the experimental parameter space available to $\mu$SR and forms the basis of the pulsed methods discussed in the following section.}

\subsection{Active manipulation \add{and pump--probe muon spectroscopy}\remove{of the muon's quantum state}}
\label{sec:active_manipulation}

\begin{figure}
    \centering
    \includegraphics[width=1\linewidth]{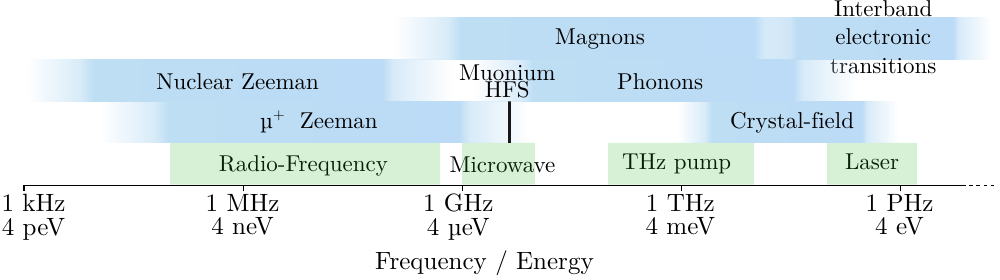}
    \caption{Excitation frequency and energy scales from kHz to PHz, illustrating representative muon–nuclear, muonium hyperfine structure (HFS), collective, and electronic transitions and the corresponding experimental electromagnetic regimes from Radio-Frequency (RF) \cite{scheuermann1997radio,cottrell1998radio,kreitzman1994microwave,lord2009microwave} to optical excitation \cite{yokoyama2013future,yokoyama2016new,yokoyama2017photoexcited} under typical laboratory conditions.}
    \label{fig:EM-excitation}
\end{figure}

If the conventional view of \musr is that the muon spin evolves under the Hamiltonian of its local environment, then active spin manipulation represents the next conceptual step: the experimenter deliberately adds a time-dependent term to the Hamiltonian \cite{kreitzman1991rf,scheuermann1997radio,cottrell1998radio,clayden2012spin}. Rather than only observing the natural evolution of the implanted muon, one can use radio-frequency, microwave, or optical excitation to drive selected transitions, perturb particular degrees of freedom, and watch how the muon responds. In this sense, the muon is not only a quantum probe embedded in a quantum system, but also a handle through which the local quantum system can be manipulated.

This way of thinking is particularly natural when considering small coupled spin clusters. The F--$\mu$ example in Figure~\ref{fig:muon_fluorine_polarisation} shows that even a minimal muon--nuclear system has a well-defined quantum level structure \cite{brewer1986observation,wilkinson2020information,Billington2022radio,chatterjee2025vector} and a polarisation function that depends sensitively on the spin Hamiltonian. The difference between the quantum spin-$\frac{1}{2}$ calculation and a classical vector-spin description highlights why the language of energy levels, eigenstates, and driven transitions is not merely decorative: as the muon polarisation reflects coherent evolution under a local Hamiltonian, applying an oscillatory field at the appropriate frequency opens up the possibility of addressing specific transitions within that Hamiltonian.

The driven Hamiltonian may be written schematically as \cite{clayden2012spin, Billington2022radio, chatterjee2025vector}
\begin{equation}
\mathcal{H}(t)=\mathcal{H}_{0}+\mathcal{H}_{\rm drive}(t),
\end{equation}
where $\mathcal{H}_{0}$ contains the static local interactions of the muon 
(examples are given in Table~\ref{hamiltonians}) and $\mathcal{H}_{\rm drive}(t)$ represents the applied electromagnetic excitation. The experimental question then changes subtly: instead of only asking how the muon's spin evolves due to its 
environment, one asks how the coupled muon--matter system responds when particular transitions or modes are driven. This is the same basic logic that underpins magnetic resonance more generally, but with the distinctive advantage that in \musr the spin-polarised probe is implanted locally and read out directly in the time domain through the muon decay asymmetry.

The range of possible excitations is summarised schematically in Figure~\ref{fig:EM-excitation}. Radio-frequency \musr (RF-$\mu$SR) occupies the MHz regime, where the relevant energy scales are typically muon Zeeman, nuclear Zeeman, and low-frequency hyperfine or dipolar transitions. In this regime, RF fields can be used to manipulate diamagnetic muons, drive transitions in coupled muon--nuclear spin systems, or decouple selected nuclear moments from the muon \cite{kitaoka1982muon,kreitzman1988spin,kreitzman1991rf,scheuermann1997radio,cottrell1998radio,clayden2012spin,cottrell2018new}. This is especially powerful when the loss of muon polarisation does not simply represent irreversible relaxation, but instead reflects coherent transfer of polarisation into a nearby spin environment. By applying suitable RF pulses, one can distinguish genuine decoherence from reversible dephasing or coherence exchange within the local spin cluster. Thus RF-\musr turns the muon from a passive reporter of local fields into an active participant in local quantum-state control.

Microwave excitation extends this idea to higher frequencies, particularly for muonium \cite{estle1983theory,kreitzman1994microwave,lord2009microwave,doll2025coherent}. Because muonium contains both a muon and an electron, its level structure is governed by the much larger electron gyromagnetic ratio and by the electron--muon hyperfine interaction. The relevant transitions therefore naturally move into the GHz range. Microwave-\musr can access these transitions directly, allowing coherent control of muonium states through Rabi oscillations, Ramsey-type measurements, and related pulse protocols \cite{doll2025coherent}. In this regime, the muon is part of a hydrogen-like quantum object whose spin and electronic degrees of freedom can be manipulated together. This opens a route to treating muonium not just as a chemical analogue of hydrogen, but as a controllable quantum system embedded in matter.

In contrast to RF and microwave excitation, optical excitation does not primarily drive the muon spin directly. Instead, laser-\musr uses light to perturb the electronic system of the host material \cite{yokoyama2013future,yokoyama2016new,yokoyama2017photoexcited}, for example by generating carriers, changing charge states, or initiating photoinduced chemical processes. The muon simply acts as a local time-domain probe of how those electronic excitations modify the magnetic, hyperfine, or charge-sensitive interactions at its site. This gives laser-\musr the character of a pump--probe experiment: the laser prepares or perturbs the material, and the muon reports on the local microscopic response.

Looking further ahead, THz-pump \musr could \add{probe long-lived changes in
the local magnetic, hyperfine, or screening environment produced by
collective excitations}\remove{provide an important intermediate regime between spin manipulation and optical excitation} \cite{yokoyama2017photoexcited,kubacka2014large,matsunaga2014light}. \add{Where a nonequilibrium excitation population or associated metastable
state persists into the nanosecond-to-microsecond regime, its relaxation
could be followed as a function of pump--probe delay. This approach would
probe the longer-lived response produced by the excitation, rather than
the intrinsic picosecond lifetime of the coherently driven THz mode.}\remove{THz photons have energies suited to low-energy collective excitations such as phonons, magnons, crystal-field transitions, or superconducting order-parameter dynamics. A THz pulse would therefore not simply heat the sample or create high-energy electronic excitations, but could selectively drive collective modes that are strongly coupled to the low-energy physics of the material. The muon would then measure how these driven modes modify local magnetic fields, spin fluctuations, charge distributions, or superconducting screening. Although this remains a future direction rather than a routine capability, it illustrates the broader opportunity: \musr can evolve from observing equilibrium quantum dynamics to probing materials while selected parts of their Hamiltonian are actively driven.}

The common theme across RF, microwave, optical, and possible THz-pump \musr is \remove{that the technique becomes explicitly Hamiltonian engineering on the timescale of the muon lifetime. One begins with a local quantum probe whose polarisation encodes the interactions around it, then applies an external perturbation to change either the state of the muon, the state of the surrounding material, or the coupling between the two. This perspective is closely aligned with the wider development of quantum materials and quantum technologies, where understanding a system increasingly means not only measuring its equilibrium properties, but also controlling its coherent and dissipative dynamics
\cite{Manzano2020,Gaidash2025,Rezvani2021}. For \musr, active manipulation therefore provides a natural future direction: it takes the quantum nature of the muon seriously and uses it not only to read out microscopic physics, but to interrogate how that physics changes when the local quantum system is driven.}\add{the combination of a controlled perturbation with local 
time-domain readout. RF and microwave fields can act directly on local 
muon-containing spin states, whereas optical and THz excitation primarily 
prepare the host material and the muon follows its subsequent relaxation. 
These approaches therefore extend \musr from equilibrium characterisation 
to time-resolved measurements of how local states and excitation 
populations evolve after excitation.}

\section{Advancing knowledge of chemical composition}
\label{negativemuon}

\subsection{Experiments with negative muons ($\mu^-$): explaining bulk phenomena with quantum processes}

\add{Although different to the approach described above with $\mu^+$, the use of negative muons ($\mu^-$) has some distinct advantages to providing knowledge on the composition, or changes in composition across a wide range of materials. This can then be used to relate back to the structure-property relationships that researchers are commonly trying to pin down.}

\begin{figure}[h]
    \centering
    \includegraphics[width=1.0\linewidth]{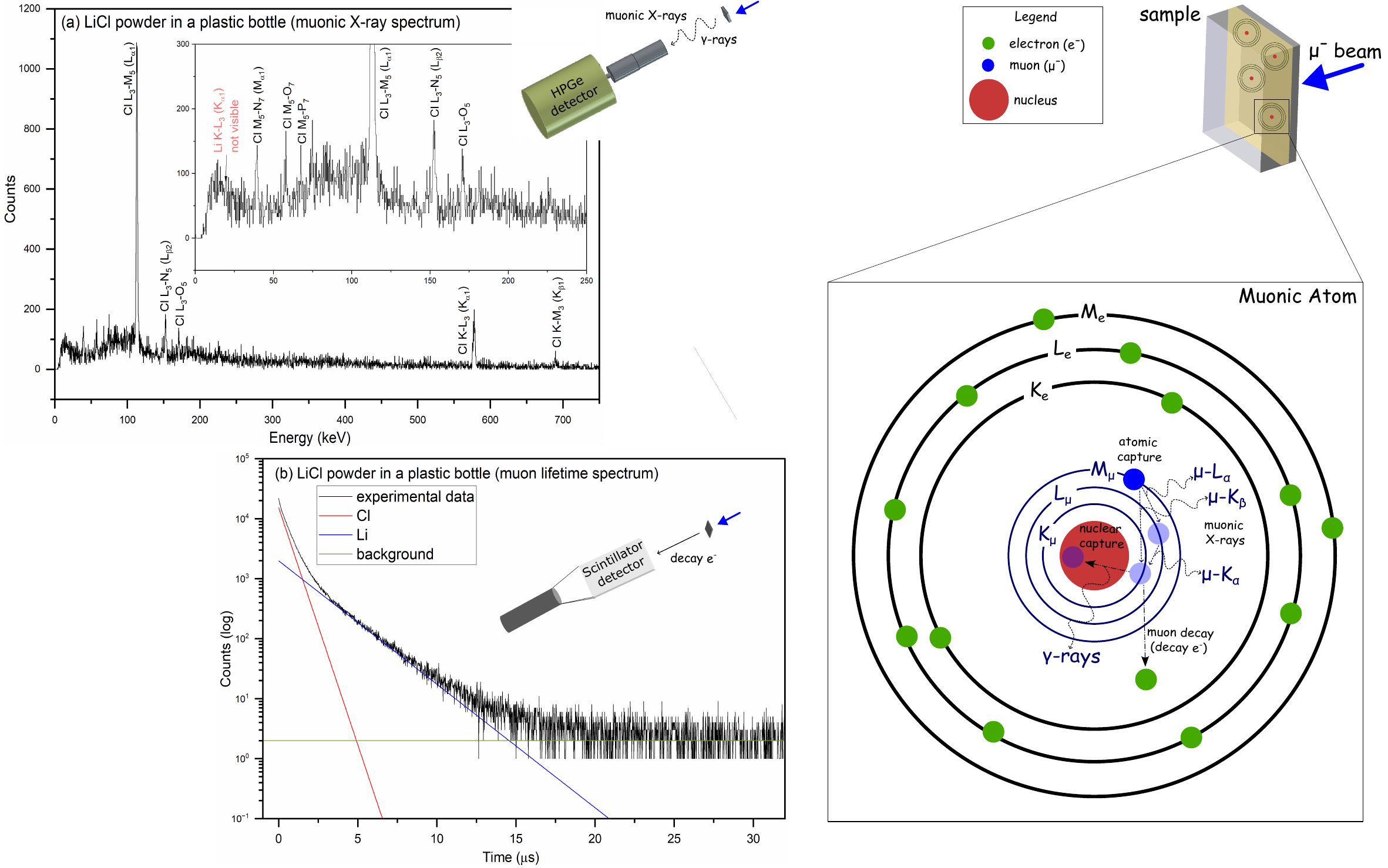}
    \caption{Pictorial representation of the different quantum processes that take place in a $\mu^-$ experiment once a sample is bombarded with a $\mu^-$ beam. The electronic and muonic orbitals in the muonic atom have been denoted with subscripts e and $\mu$, respectively. A lithium chloride (LiCl) powder sample, packed in a plastic bottle with dimensions \add{$\sim7 \times 7 \times 12$~cm} \remove{$5 \times 5 \times 10$~cm} was used to obtain the experimental (a) muonic X-ray spectrum using the MuX spectrometer, an array of High Purity Germanium (HPGe) semiconductor detectors and (b) muon lifetime spectrum using the ARGUS spectrometer, an array of plastic scintillator detectors, at the ISIS Neutron and Muon Source. Both the IUPAC (K-L$_3$, for example) and Siegbahn (K$_{\alpha1}$, for example) notations, where possible, have been used to denote the characteristic muonic X-ray peaks in (a).}
    \label{fig:licl_negative_muon_experiment}
\end{figure}

Negative muons, unlike positive muons ($\mu^+$), get captured by the atom of the different elements present inside a given sample, forming a muonic atom.
\add{Figure~\ref{fig:licl_negative_muon_experiment} gives a pictorial description of the quantum processes involved during the interaction of a $\mu^-$ beam with a sample.}
\add{The muonic atom, thus formed, subsequently emits}\remove{This leads to the emission of} muonic X-rays, the experimental realisation of 
which has led to the development of a non-destructive \add{elemental analysis} technique called muonic X-ray Emission Spectroscopy ($\mu$-XES)\footnote{Unfortunately, a common acronym for this technique has still not been agreed upon and is also known by other names, namely, muonic X-ray Analysis (MXA)~\cite{Reidy1978}, Muonic Atom X-Ray Spectroscopy (MAXRS)~\cite{Clemenza2019}, and Muon Induced X-ray Emission (MIXE)\cite{Biswas2022}.}, which is very similar to X-Ray Fluorescence (XRF)~\cite{Potts1992}.
This technique, developed back in the 1970s~\cite{Taylor1973, Reidy1978, Kohler1981} has seen a recent resurgence in usage with increasing beam intensities and dedicated experimental setup in large-scale accelerator facilities~\cite{Hillier2016, Terada2017, Biswas2022}. 
\remove{Figure~\ref{fig:licl_negative_muon_experiment} gives a pictorial description of the quantum processes involved during the interaction of a $\mu^-$ beam with a sample.}
The key difference between XRF and the $\mu$-XES technique is that the energies of muonic X-rays are much larger than the corresponding electronic X-rays, which makes it possible for muonic X-rays to penetrate from deep inside the sample and reach the detectors, with \add{relatively lower} \remove{minimum} self-absorption inside the sample.
\add{For example, the electronic K$_{\alpha}$ X-ray of Cu is at $\sim$8~keV\cite{Mendenhall2017}, while the \textbf{muonic} K$_{\alpha}$ X-ray of Cu is at $\sim$1514~keV\cite{Biswas2022}. 
Due to this difference, one is able to probe only a few \textmu m inside Cu with XRF, while with muonic X-rays, one can probe as deep as a few millimetres in Cu.}
In addition, by changing the energy of the $\mu^-$ beam, one can control the probing depths inside the sample.
\remove{For example, the electronic K$_{\alpha}$ X-ray of Cu is at $\sim$8~keV\cite{Mendenhall2017}, while the \textbf{muonic} K$_{\alpha}$ X-ray of Cu is at $\sim$1514~keV\cite{Biswas2022}. 
Due to this difference, one is able to probe only a few \textmu m inside Cu with XRF, while with muonic X-rays, one can probe as deep as a few mm in Cu.}
\add{For example, if we have a sample consisting of a sandwich of Fe and Cu foils with thicknesses 100 and 300~${\mu}$m respectively, with the Fe layer facing the beam, a $\mu^-$ beam at 20~MeV/$c$ would cause all muons to stop inside Fe layer, leading to emission of muonic X-rays from Fe. 
However, when a $\mu^-$ beam at 30 MeV/$c$ is bombarded on the same sample, all the muons would stop in the Cu layer giving muonic X-rays of Cu.}

\remove{The} \add {Due to the aforementioned properties, the} $\mu$-XES technique has been used recently to determine the elemental composition in a non-destructive manner in archaeological artefacts~\cite{Biswas2023, Hofmann2023, Green2022, Hampshire2019, ninomiya2015elemental}. 
This technique has also been utilised to study battery materials, especially Li-ion batteries (LIBs)~\cite{Querel2025, Umegaki2025}, the analysis for which is challenging due to the overlapping peaks of muonic K$_{\alpha}$ X-rays of Li (\add{18.7~keV}) and muonic L$_{\beta}$ X-rays of C (\add{18.8~keV}).
In addition, the detection of light elements, like Li, in thicker materials using this technique has proved to be challenging due to the attenuation/self-absorption of the low-energy muonic X-rays in these thicker samples, as shown in Fig.~\ref{fig:licl_negative_muon_experiment}(a).
Also, depending on the setup/muon facility, the detection limit for elements, using this technique is $\sim$0.5-1~wt$\%$.

\add{After the muonic X-ray cascade process, the muon in the muonic atom undergoes one of two processes: (i) it can be captured by the nucleus which eventually leads to emission of gamma rays or (ii) it decays to electrons and hence has a definite lifetime.}
Very recently, there have been attempts to use an alternative technique using $\mu^-$ beams, the mean lifetime of $\mu^-$, to also determine the elemental composition, specifically looking into carbon content in steel samples and ancient Japanese steel swords~\cite{Ninomiya2024, kiyanagi2026metallurgical}.
These previous publications, claim a detection limit of $\sim$0.2 wt$\%$, using the muon lifetime technique. 
The lifetime of $\mu^-$ is different in different elements; the lifetime decreases with increasing atomic number due to competition with the nuclear capture process~\cite{Yama1975, Suzuki1987, Measday2001}. 
For example, the lifetime of  $\mu^-$ in vacuum is $\sim$2.2~${\mu}$s, while that in Cu is $\sim$160~ns~\cite{Suzuki1987}.
An example muon lifetime spectrum has been shown in Fig.~\ref{fig:licl_negative_muon_experiment}(b).
The Li muonic X-rays, not visible in Fig.~\ref{fig:licl_negative_muon_experiment}(a) are easily identified in  Fig.~\ref{fig:licl_negative_muon_experiment}(b).
Owing to the challenges in the quantification of Li in LIBs and thicker samples using the muonic X-ray technique (mentioned in the previous paragraph), the usage of the muon lifetime technique to LIBs is being investigated~\cite{Biswas202x}.

It is the combination of the two aforementioned techniques \add{(more details on these two techniques can be found in the Supporting Information)} that allows one to map the elemental distribution across a sample. 
Many technologically relevant systems rely on the motion or movement of ions, and can comprise a range of different elements spanning from heavy to light. 
Biological processes, energy storage materials, and some industrial materials tend to focus on the motion/quantification of lighter elements and being able to depth profile heavy and light element composition using the two negative muon techniques allows one to build up a complete picture.

\section{Final Thoughts}

We have shown that muon spin spectroscopy is most useful when we stop 
thinking of it as a niche technique and start treating it as a genuinely distinctive way 
of looking at quantum processes within matter. This reframing spans a wide range of materials 
chemistry and physics; not just ultra low temperature quantum states. The muon does 
not passively sample a material, instead it is part of the local quantum environment, and 
its spin evolution gives direct access to the interactions that are important on the 
scales where a lot of the interesting physics and chemistry sit. That is what makes \musr so valuable: it is not just sensitive, but it is sensitive to the right things, in the 
somewhat awkward middle ground between local behaviour and bulk response where 
many important materials problems live.

That also helps explain why the technique keeps showing up in such a wide range 
of areas. Whether the question is about decoherence, diffusion, charge rearrangement,
defects, or hydrogen-like chemistry, the common thread is that the important behaviour 
is local, dynamic, and often difficult to get to the bottom of using other methods and techniques. \musr does not 
replace those techniques; its strength is that it adds something genuinely different 
and in some cases, special. For most problems, it is precisely the fact that the muon 
is local and quantum mechanical that makes it useful. 

So, the broader point is not that muons need better marketing as a more conventional 
probe. It is really the opposite. The future success of the field lies in leaning harder into 
what makes the technique unusual: the quantum nature of the muon, the flexibility of the 
states it forms in matter, and the fact that it can connect microscopic interactions to real 
material properties in ways that are hard to do otherwise. If that is where things are 
heading, then the muon is no longer the oddity in the experimental toolbox; it 
is becoming one of the more incisive ways of asking what a material is actually doing 
and how it does it. Rabi may have asked “Who ordered that?”, but at this point it 
is fair to say the materials community probably should have.

\begin{acknowledgement}

The authors would like to thank all those that came before us.

\end{acknowledgement}

\begin{suppinfo}
Brief summaries of the different techniques using $\mu^+$ and $\mu^-$\add{, together with a technical discussion of the relationship between local $\mu$SR relaxation and wavevector-dependent correlations,} are provided within the Supporting Information.

\end{suppinfo}


\bibliography{MuonsFuturePerspective.bib}

\end{document}